\documentclass[prl,showpacs,floatfix,amsmath,amsfonts,twocolumn]{revtex4}  
\usepackage{graphics}
\usepackage{epsfig}
\usepackage{color}

\newcommand {\tmu}{\tilde{\mu}}

\newcommand{\p}{{\cal P}}
\newcommand{\PT}{{\cal PT}}
\newcommand{\T}{{\cal T}}
\newcommand{\cA}{{\cal A}}

\newcommand{\bPsi}{\mathbf{\Psi }}
\newcommand{\bPhi}{\mathbf{\Phi }}
\newcommand{\bpsi}{ \mbox{\boldmath$\psi$\unboldmath}  }
\newcommand{\bphi}{ \mbox{\boldmath$\phi$\unboldmath}  }

\newcommand{\bu}{{\bf u}}

\newcommand{\bA}{{\bf A}}

\newcommand{\kmin}{k_{\rm min}}

\begin{document}

\title{Solitons in Bose-Einstein Condensates with Helicoidal Spin-Orbit Coupling }

\author{
	Yaroslav V. Kartashov$^{1,2,3}$ and Vladimir V. Konotop$^{4}$
	}

\affiliation{ 
$^{1}$ICFO-Institut de Ciencies Fotoniques, The Barcelona Institute
of Science and Technology, 08860 Castelldefels (Barcelona), Spain 
\\
$^{2}$Institute of Spectroscopy, Russian Academy of Sciences, Troitsk,
Moscow Region, 142190, Russia
\\
$^{3}$Department of Physics, University of Bath, BA2 7AY, Bath, United Kingdom 
\\
$^{4}$Centro de F\'isica Te\'orica e Computacional and Departamento de F\'isica, Faculdade de Ci\^encias,  Universidade de Lisboa, Campo Grande 2, Edif\'icio C8, Lisboa 1749-016, Portugal
}

\date{\today}

\begin{abstract}

We report on the existence and stability of freely moving solitons in a spatially inhomogeneous Bose-Einstein condensate with helicoidal spin-orbit (SO) coupling. In spite of the periodically varying parameters, the system allows for existence of stable propagating solitons. Such states are found in the rotating frame, where the helicoidal SO  coupling is reduced to a homogeneous one. In the absence of the Zeeman splitting the coupled Gross-Pitaevskii equations describing localized states feature many properties of the integrable systems. In particular, four-parametric families of solitons can be obtained in the exact form. Such solitons interact  elastically. Zeeman splitting still allows for the existence of two families of moving solitons, but makes collisions of solitons inelastic.
 
\end{abstract}

\maketitle

{\color{black} When parameters of a continuous medium vary periodically the translational invariance is broken and  nonlinear localized excitations cannot propagate freely}. This is at variance with the linear systems, where  the Bloch theory allows for quantum particles or waves to move without backscattering. The well-known examples are  electrons in solids, atoms in optical latices, electromagnetic waves in photonic crystals, and many others. Physical understanding of the interplay between the nonlinearity and periodicity is simple. Since the nonlinear excitations are localized, in the presence of a periodic potential their energy depends on the spatial location of the wavepacket. 
Respectively, a steady forward motion is impossible because of the potential barriers {\color{black} causing radiative losses of moving localized wavepackets.}. These facts are well documented, both theoretically and experimentally, in the physics of Bose-Einstein condensates (BECs) ~\cite{BEC-lattices} and in nonlinear optics~\cite{photonic}.  {\color{black} 
	Several approaches to obtaining moving solitons in periodic media were suggested. Radiation is reduced for sufficiently wide and small-amplitude solitons in linear lattices (described using envelope function approach in photonics~\cite{Sterke} or effective mass approximation in the meanfield theory~\cite{KonSal}), as well as in nonlinear lattices~\cite{nonlin_latt}. Mobility of strongly localized solitons can be enhanced in lattices with saturable, quadratic or nonlocal nonlinearities, as well as in materials with competing linear and nonlinear lattices (see \cite{Kart_rev} for a review). However, in all these models radiation is not arrested completely: it becomes detectable at large propagation distances.}

In this Letter we show that {\color{black} freely moving nonlinear waves can exist if a system}  obeys special symmetries. {\color{black} In contrast to all previous studies, the solitons reported here do not radiate and propagate over infinitely long distances for any peak amplitude or any ratio of soliton width to system period.} As the case example we consider a spin-orbit (SO) coupled BEC which is well accessible in laboratories~\cite{Nature,Engels} and represents a versatile tool for study of the nonlinear physics of synthetic fields~\cite{experiments} and gauge potentials~\cite{Ruseckas,tunable_SO}. 
We consider a SO-BEC described by the Hamiltonian, whose linear part reads $\displaystyle{H_{\rm lin}=  \left[p +\alpha \bA(x) \right]^2/2+{\Delta}\sigma_3/2}$, where  $p=-i\partial/\partial x$ is the linear momentum operator, $\bA(x)$ is the spatially varying gauge potential, $\alpha$ is the potential amplitude, $\Delta$ is the Zeeman splitting, and we use the units with $m=\hbar=1$ as well as notations $\sigma_{1,2,3}$ for the Pauli matrices. The SO coupling, whose strength is experimentally tunable {\color{black} using different techniques}~\cite{tunable_SO} {\color{black} (see also~\cite{tunable_theor}, for theoretical discussion)}, is considered of the helicoidal shape with the period $\pi/\kappa$, i.e. $\bA(x)=\boldsymbol{\sigma}\mathbf{n}(x)$, where $\mathbf{n}(x)=\left(\cos(2\kappa x),\sin(2\kappa x),0\right)$   and  $\boldsymbol{\sigma}=\left(\sigma_1,\sigma_2,\sigma_3\right)$. Inter- and intra-species interactions are attractive and equal. Then the spinor order parameter $\boldsymbol{\Psi}=\left(\Psi_1,\Psi_2\right)^T$ solves   the vector Gross-Pitaevskii (GP) equation 
\begin{equation}
\label{GPE}
i\frac{\partial \bPsi}{\partial t}=\frac{1}{2}\left(\frac 1i \frac{\partial}{\partial x}+\alpha \bA(x)\right)^2\bPsi+\frac{\Delta}{2}\sigma_3\bPsi-\left(\bPsi^\dag\bPsi\right)\bPsi
\end{equation} 

Previously periodic SO coupling was considered only in (intrinsically linear) solid state applications~\cite{PeriodSO_solid}. Helicoidal gauge potential can also be created in optical systems, where it arises in description of light propagation in helical waveguide arrays~\cite{Szameit}. We also mention a study~\cite{Caetano} of spin currents due to SO coupling in a nonlinear model of a DNA helicoidal molecule. 
Solitons in uniform SO-BECs were investigated in~\cite{AFKP}. {\color{black} Immobile excitations pinned to lattice sites} were reported for SO-BECs with optical~\cite{OL} and Zeeman~\cite{KKA} lattices. Dynamical effects, like temporal management of solitons in deep optical lattices~\cite{SAGT}, and interaction of solitons in BECs with localized SO coupling~\cite{KKZ}, have been reported, as well. {\color{black} 
	Soliton motion in such systems is accompanied by radiative losses.}
 
The helicoidal structure of the vector potential 
introduces the point translational symmetry (the shift by the period $\pi/\kappa$). The central observation of this Letter is that for the chosen gauge field $\bA(x)$, switching to the rotating frame using gauge transformation
$
\bPsi=e^{-i(\alpha^2+\kappa^2)t/2}e^{-i\sigma_3\kappa x}\bpsi
$
changes the point translational symmetry of Eq.~(\ref{GPE}) into continuous translational symmetry of the transformed equation. The new spinor $\bpsi$ solves the GP equation with constant coefficients:
 \begin{equation}
 \label{GPE_transformed}
%
i\bpsi_t=-\frac{1}{2}\bpsi_{xx}-i\left(\alpha\sigma_1 +\kappa\sigma_3\right) \bpsi_x 
  +\frac{\Delta\sigma_3}{2} \bpsi-(\bpsi^\dag\bpsi)\bpsi.
 \end{equation}

\begin{figure}[t]
\includegraphics[width=1\columnwidth]{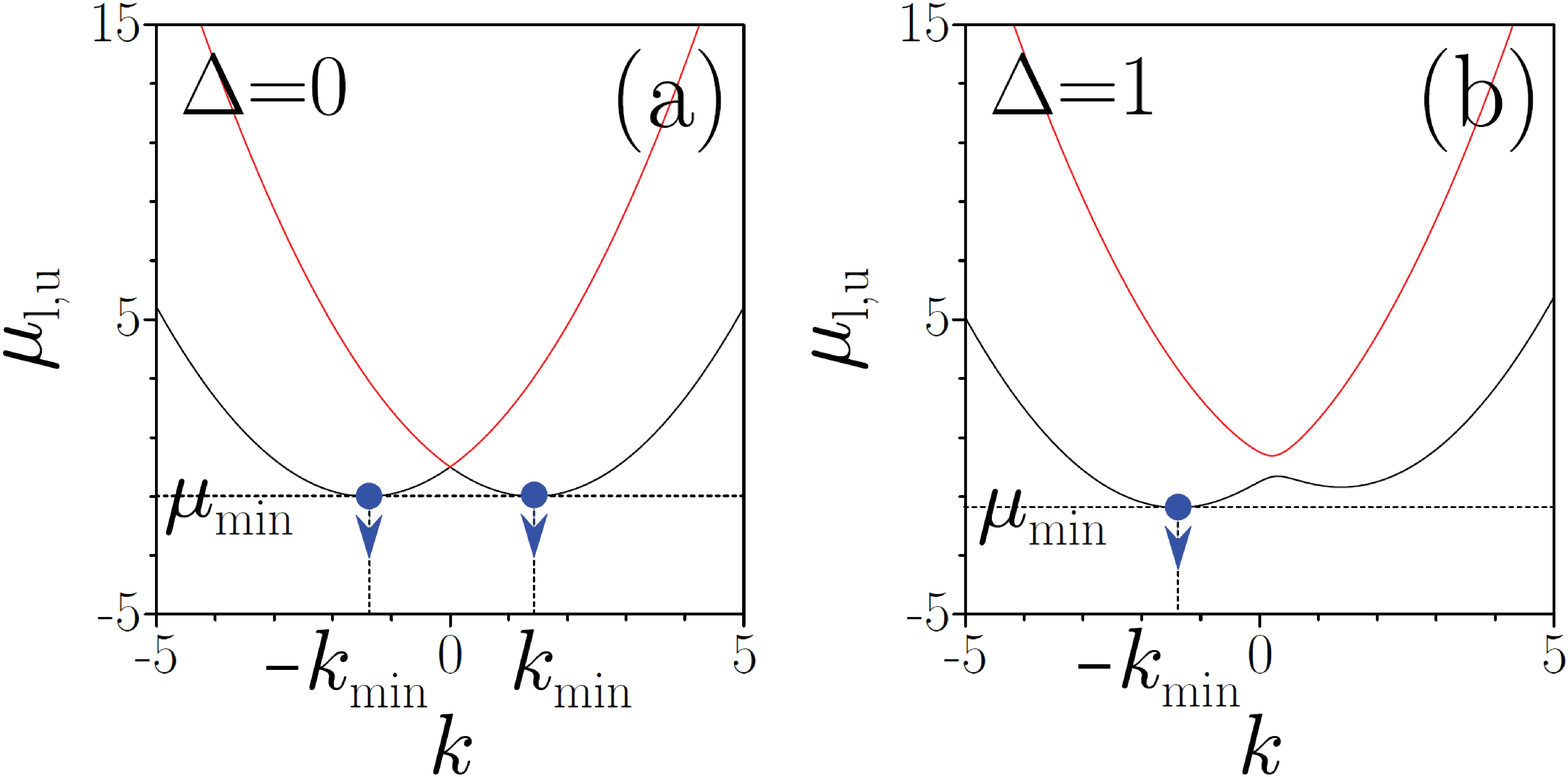}
 \caption{Dispersion curves $\mu_{\rm l}$ (blue lines) and $\mu_{\rm u}$ (red lines) for $\alpha=\kappa=1$; "l" and "u" stand for lower and upper branches. 
 	Arrows show bifurcation points for quiescent nonlinear modes.}
\label{fig:one}
\end{figure}

Thus although the original system (\ref{GPE}) had periodically varying parameters, the field $\bpsi$ obeys the equation with $x$-independent coefficients and with the linear spectrum typical for unequal Rasba~\cite{Rashba} and Dresslhaus~\cite{Dresselhaus} couplings, i.e. $\mu_{\rm l,u} ={k^2}/{2}\pm \sqrt{\alpha^2k^2+(\kappa k-\Delta/2)^2}$, shown in Fig.~\ref{fig:one}.  Equation~(\ref{GPE_transformed}) reveals also an important role of the Zeeman splitting. At $\Delta=0$ the system obeys the  $\PT$-symmetry with the conventional spatial inversion  $\p\bpsi(x,t)=\bpsi(-x,t)$ and (integer spin) time reversal $\T$:  $\PT\bpsi(x,t)=\bpsi^*(-x,-t)$. On the other hand, Eq.~(\ref{GPE_transformed})  obeys also the (half-integer spin~\cite{Messiah}) time-reversal symmetry $\T_F=\sigma_2 \T$, as well as the symmetry with respect to transformation $\cA=\sigma_2\p$.  These operators $\{\PT, \T_F, \cA\}$ completed by the identity operator constitute the Klein-four group. Thus a localized solution (if any) either possesses all three symmetries (then it is highly symmetric) or obeys only one of them~\cite{LKK}. However, if $\Delta>0$ only $\PT$ symmetry remains ($\T_F$ and $\cA$ being broken).  
 
\paragraph{Solitons at zero Zeeman splitting.} At $\Delta=0$ the lower branch of the dispersion relation obeys two equal minima  $ \mu_{\rm min}=\min_{k}\mu_{\rm l}=-\kmin/2$ achieved at $k=\pm k_{\rm min}$, where $k_{\rm min}=\sqrt{\alpha^{2}+\kappa^{2}}$. {\color{black} Quiescent solitons can only exist in the parameter domain where propagation of linear modes is prohibited,  i.e. } at $\mu<\mu_{\rm min}$ [see Fig.~\ref{fig:one} (a)]; they belong to the families bifurcating from the linear spectrum. It turns out, that the system admits also four-parametric families of {\em moving} solutions which can be obtained analytically. 
 
 Indeed, introducing the momentum $Q=-i {\partial}/{\partial x}+\alpha \bA(x)$ we rewrite Eq.~(\ref{GPE}) at $\Delta=0$ in the form: $i\bPsi_t= Q^2\bPsi/2-(\bPsi^\dag\bPsi)\bPsi$. Soliton solutions of this equation can be constructed using the basis of the eigenfunctions $\bPhi_\pm$ of the linear operator $Q$: $Q\bPhi=q\bPhi$. 
 This last problem is solved giving two eigenvalues $q_\pm=k\pm k_{\rm min}$ with the respective eigenmodes
\begin{equation}
\label{sol+}
 \bPhi_\pm=e^{ikx}  \left(
	\!\!\begin{array}{c}
 -e^{-i\kappa x}\sin\nu_\pm 
	\\ 
	e^{i\kappa x}\cos\nu_\pm  
	\end{array}
	\!\!\right)\!, \,\,\, \nu_\pm=\frac 12 \arctan \frac{\kappa}{\alpha}\mp\frac{\pi}{4},
 \end{equation}
which are orthonormal: $\bPhi_\pm^\dag\bPhi_\pm= 1$, $\bPhi_\mp^\dag\bPhi_\pm=0$. Notice, that the quantities  $q_\pm^2/2$  define detuning of the chemical potentials from the bottom of the linear spectrum $\tmu_{\rm min}$. 

Next we look for a solution of (\ref{GPE}) with $\Delta=0$ in the form $\bPsi=u_+e^{-iq_+ x}\bPhi_++u_-e^{-iq_- x}\bPhi_- $, where  $u_{\pm}(t,x)$ are complex functions. It is straightforward to obtain that $u_{\pm}$ solve the Manakov system~\cite{Manakov}
\begin{eqnarray}
\label{Manakov1}
i\bu_t=-(1/2) \bu_{xx}-(\bu^\dagger\bu)\bu , \qquad \bu=(u_+,u_-)^T.
\end{eqnarray} 
A diversity of solutions of (\ref{Manakov1}) can be constructed using the methods of exactly integrable systems. We mention only the simplest soliton solution with $u_+=u_-$: 
\begin{eqnarray}
\label{stripe_psi}
\bPsi_{\rm sol}^{(\pm)}  
=\frac{\eta e^{i\left(vx-(v^2-\eta^{2}-\kmin^{2})t/2\right)}}{{2}^{1/2}\cosh(\eta(x-vt))} 
\nonumber \\
\times
\left(\!\!
\begin{array}{c}
\left(- e^{-i\kmin x} \sin\nu_+ 
-
  e^{i\kmin x}\sin\nu_- \right)e^{-i\kappa x} 
\\[1mm]
\left( e^{-i\kmin x}\cos\nu_+  
+    e^{i\kmin x}\cos\nu_-  \right)e^{i\kappa x} 
\end{array}
\!\!\right), 
\end{eqnarray}
where $v$ and $\eta$ describe the velocity and the amplitude of the soliton. In the conventional terminology~\cite{AFKP}, the solutions $\bPsi_{\rm sol}^{(\pm)}$ are {\em stripe} solitons since at $v=0$ they bifurcate from two minima $\mu_{\rm min}$ at the points $\pm k_{min}$ shown with blue arrows in Fig.~\ref{fig:one} (a).
Meantime, Eq.~(\ref{Manakov1}) is SU(2) invariant. This means that if $\bu$ is a solution, then $\tilde{\bf u}=S\bu$, where $S=\left(\begin{array}{cc}
\gamma & \delta \\ -\delta^* & \gamma^* \end{array}\right)$ and the parameters $\gamma$ and $\delta$ are linked by $|\delta|^2+|\gamma|^2=1$, is a solution too. 
This transformation erases the difference between stripe and {\em conventional} solitons, the latter understood as bifurcating either at $\kmin$ or at  $-\kmin$ and corresponding to  either $\tilde{u}_-=0$ or $\tilde{u}_+=0$, since both are related by the simple rotation $S$ with properly selected elements. Thus, all solitons appear as particular members of a family parametrized by $v$, $\eta$, $\delta$, and $\gamma$. 
 
Although the use of specific ansatz above resulted in an exactly integrable system (\ref{Manakov1}), it does not ensure integrability of the original model (\ref{GPE}). 
Therefore we studied stability and interactions of solitons at $\delta=0$ in the frames of Eq.~(\ref{GPE}).
 We observed both stability of solitons and their elastic interactions, which are characteristic for solitons in integrable systems (not shown here).

\paragraph{Solitons at non-zero Zeeman splitting.} As mentioned above, at $\Delta>0$ only $\PT$-symmetry remains unbroken. Now the branch $\mu_{\rm l}(k)$ acquires a non-degenerate minimum at $-k_{\rm min}$, so that quiescent solitons are expected to bifurcate from $\mu_{\rm l}(-k_{\rm min})$ [Fig.~\ref{fig:one}(b)]. It turns out that, at least for small and even moderate SO coupling, one can find pairs of different solitons, such that solutions within a pair feature equal chemical potentials and numbers of atoms, the latter defined as $N=\int_{-\infty}^{\infty}\bpsi^\dag\bpsi\, dx$. To show this we concentrate on quiescent solutions $\bpsi(t,x)=e^{-i\mu t}\bphi (x)$ of the system (\ref{GPE_transformed}).  If, in the absence of the SO coupling ($\alpha=0$) such a solution $\bpsi (x)$ is found, then there exists also another solution $e^{-i\mu t}\sigma_3\bphi(x)$ corresponding to the same $\mu$ and $N$. The families of these unperturbed (by $\alpha$) solutions coincide and are described by the function $\mu(N)$. 
Next, one can show~\cite{supplemental} that a small $\alpha$ ($\alpha\ll 1$) results in the splitting of the chemical potential: $\mu(N)\pm\alpha \mu_1(N)$, where $\mu_1(N)$ is also a function of $N$. In other words, the solutions $\bphi$ and $\sigma_3\bphi$ acquire opposite shifts of the chemical potentials and thus belong to different families. The (nonlinear) eigenvalue $\mu+\alpha |\mu_1|$ corresponds to the family bifurcating from the linear spectrum at $-k_{\rm min}$, and hence having the linear limit $N_-\to 0$ at $\mu\to\mu_{\rm min}$ (we call it "$-$" family). The second family ("$+$" family) with $\mu-\alpha |\mu_1|$ is "detached" from the linear eigenmode $k_{\rm min}$ and at $\Delta>0$ features the excitation threshold, i.e. $\min N_+=N_{\rm th}>0$. For a given $\mu$ and $\alpha>0$ we have $N_-<N_+$.  

\begin{figure}
	\includegraphics[width=1\columnwidth]{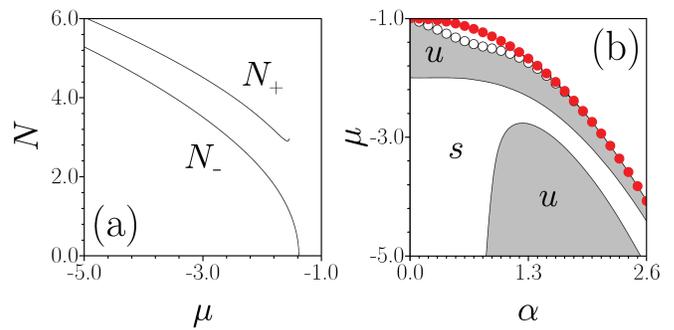}
	\caption{(a) Families of the quiescent  solitons at $\alpha=\Delta=\kappa=1$. (b) Domains of existence and stability for quiescent  "$+$" solitons  on the plane $(\alpha,\mu)$. Line with open circles shows upper cutoff for the chemical potential for "$+$" family. Line with red circles shows bottom of continuous spectrum (and thus the upper cutoff for the "$-$" family). Solitons are stable in the white domains and unstable in the gray domains.   }
	\label{fig:two}
\end{figure}

The described properties are observed also for moderate $\alpha$, as illustrated in Fig.~\ref{fig:two} (a) by the curves $N_\pm(\mu)$ and in Fig.~\ref{fig:two} (b)  by the domains of existence on the plain ($\alpha,\mu)$. In the last panel the family $N_-$ exists below the bottom of the linear spectrum $\mu_{\rm min}$ (the line with filled red circles) while the upper cut-off of the chemical potential for $N_+$ family is shown by the line with open circles. Both cutoffs decrease and gradually merge as the SO coupling increases. 
The solitons of the "$-$" family, bifurcating from the linear spectrum are stable in the entire domain of existence [i.e. below the line with red circles in Fig.~\ref{fig:two} (b)], "$+$" solitons are unstable in the region near the upper cut-off of the chemical potential and in additional domain appearing for sufficiently large SOC strength [the lower gray domain in Fig.~\ref{fig:two} (b)]. The gray region near cutoff consists of multiple alternating narrow stability and instability domains which are not resolved here.

\begin{figure}
	\includegraphics[width=1\columnwidth]{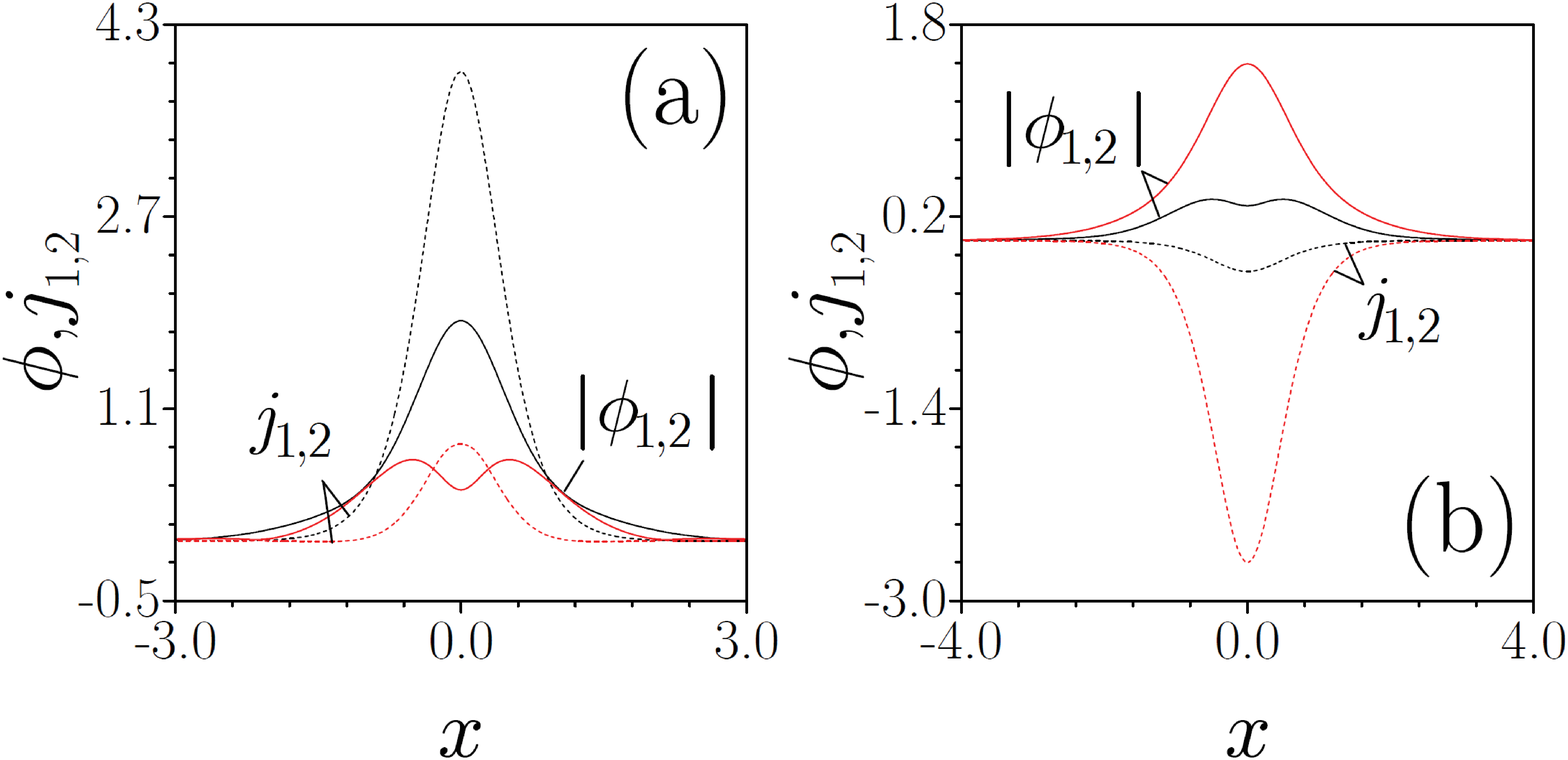}
	\caption{Field modulus (solid lines) and current (dashed lines) in quiescent solitons belonging to "$+$" (a) and "$-$" (b) families at $\mu=-2.5$ and $\alpha=\Delta=\kappa=1$. Black and red lines correspond to the first and second components.}
	\label{fig:three}
\end{figure}

Examples of the quiescent solitons for both families are shown in Fig.~\ref{fig:three}. The atomic distributions between two spinor components are imbalanced and solitons are characterized by the nonzero internal current densities in each of the components: $j_{n}=\frac{1}{2}(\psi_n^*p\psi_n-\psi_np\psi_n^*)$ [these properties were also verified in the absence of Zeeman coupling, see (\ref{stripe_psi})]. One can observe that while the higher populated states are always bell-shaped, the components with smaller number of atoms have well-pronounced density minimum in the center. Currents have single maximum and they have opposite signs for solitons of different types. In terms of the total spin projections $s_j=\frac 12 \int \bPsi^\dag\sigma_j\bPsi dx$ [$s_0$ describes also the total density distribution, while $s_3$ is the population imbalance of  the atomic states] "$+$" and "$-$" families are characterized by the opposite signs of $s_3$. 

\begin{figure}
	\includegraphics[width=\columnwidth]{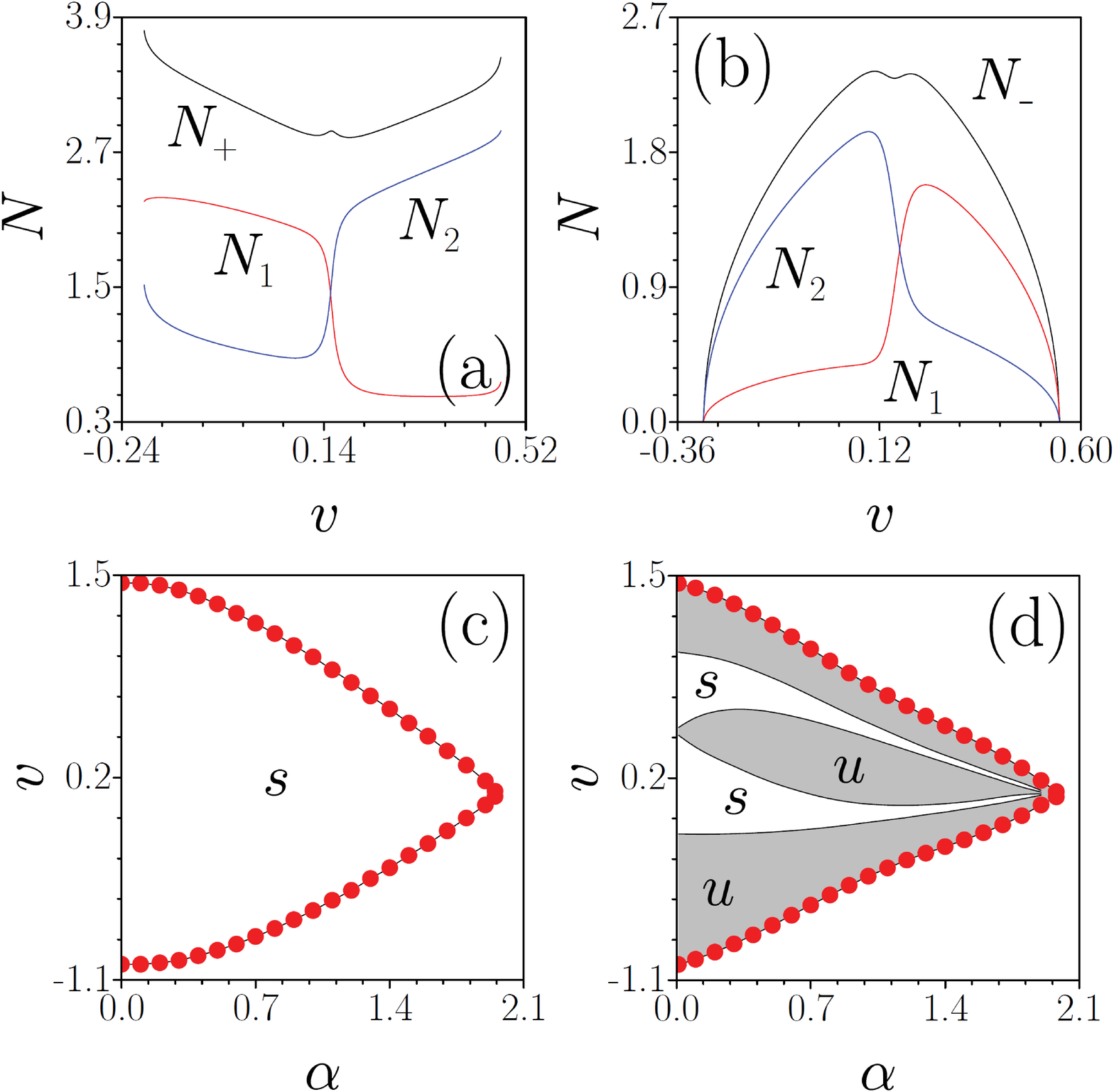}
	\caption{Numbers of atoms and their distribution between the components $N_{1,2}$ in "$+$" (a) and "$-$" (b) families {\it vs} velocity $v$ at $\mu=-2.5$, $\alpha=1.5$ and $\Delta=\kappa=1$. Domains of existence and stability for "$-$" (c) and "$+$" (d) solitons on the plane $(\alpha,v)$. Solitons are stable in white areas and unstable in gray areas. }
	\label{fig:four}
\end{figure}

In the presence of the Zeeman splitting one still can find solitons moving with a constant velocity. The dependences of the number of atoms in "$+$" and "$-$" solitons on the velocity, obtained numerically are shown in Fig.~\ref{fig:four}~(a) and~(b). Both families exist within the limited range of the velocities $v\in I_v^{\pm}=[v_{\rm min}^{\pm}, v_{\rm max}^{\pm}]$. The number of atoms in the "$-$" family vanishes at two cutoff values of velocity $v_{\rm min,max}^{-}$, while in the "$+$" family the line tangential to $N(v)$ dependence becomes vertical at the borders of the existence domain, i.e. at $v_{\rm min,max}^{+}$. In the absence of Galilean invariance change of the soliton velocity leads to variation of the ratio of atoms in the components. In Figs.~\ref{fig:four}~(a) and \ref{fig:four}~(b) we observe the inversion of the $z$-projection of the spin and  of the state populations $N_{1,2}=\int_{-\infty}^{\infty}|\psi_{1,2}|^2dx$ (because $s_3=N_1-N_2$) with increase of the velocity.

The intervals of existence $I_v^{\pm}$ strongly depend on the SO coupling: they collapse to a point as $\alpha$ grows [see Fig.~\ref{fig:four} (c) and (d)]. Here we again observe different stability properties of the "$-$" and "$+$" solitons; while the former are stable in the entire domain $I_v^-$ for a given $\alpha$ [panel (c)], the stability and instability domains for "$+$" solitons alternate with each other [panel (d)].
   
\paragraph{Spin dynamics in soliton interactions.} The existence of stable moving solitons raises the question about their interactions. In our case one can distinguish four different types of soliton collisions involving two solitons of the same family and of different families. Here we address only interactions of slow solitons propagating towards each other and concentrate on the spin dynamics.  

\begin{figure}
	\includegraphics[width=1\columnwidth]{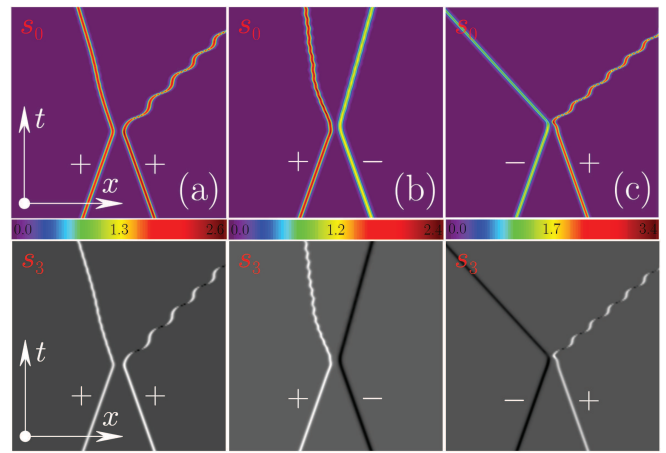}
	\caption{Collision of two solitons in the $(x,t)$ plane at $\mu=-2.5$, $\alpha=0.5$, and $\Delta=\kappa=1$. Top and bottom rows show the atomic density distributions $s_0$ and distribution of $z$-projection of the spin $s_3$, respectively.  All left (right) solitons have initial velocity $v=+0.14$ ($v=-0.14$). The types of colliding solitons ("$+$" or "$-$") are indicated on the plots. Transverse window $x\in[-30,30]$ is shown. Solitons evolve up to $t=150$. On the black-and-white panels lighter (darker) domains correspond to spin-up, $s_3>0$, (spin-down, $s_3<0$) states.  }
	\label{fig:five}
\end{figure}

Collisions of the solitons  of "$-$" family, which bifurcates from the linear spectrum, are nearly elastic and are not shown here. Meantime, interactions of either two "$+$" solitons or solitons belonging to different families reveal several inelastic effects. We describe them using the illustrative examples of Fig.~\ref{fig:five}. All shown collisions share several common features; they reveal broken $\p$ symmetry, result in spatially localized spinor states, and feature breathing character of the emergent pulses  (manifested in spin rotation and in an oscillatory trajectory).  

In all the cases, for each incident "$-$" soliton there exist emergent "$-$" soliton propagating with (slightly different) constant velocity in the opposite direction. In this sense one may speak about repulsion of solitons: no flip of the spin occurs [bottom panels in Figs.~\ref{fig:five} (b) and (c)]. The result of the interaction of "$+$" solitons depends on whether their initial velocity has the same [Fig.~\ref{fig:five} (a) and (b)] or opposite [Fig.~\ref{fig:five} (a) and (c)] direction as compared with SO coupling (i.e. with $\alpha \langle p\rangle$, where $\langle p\rangle$  is the average momentum of the initial pulse). In the former case $s_3$ keeps its sign, but its modulus undergoes variations due to spin rotation. Respectively, the projections $s_{1,2}$ are also changing (since $s_1^2+s_2^2+s_3^2=s_0^2$).  In the latter case we observe right-propagating spinor with periodically exchanged spin-up and spin-down states. The spinor rotation occurs with the spatial period $2\pi/\kappa$ (as follows from the relation between $\bPsi$ and $\bpsi$), which is very close 
to spatial scale $\approx 6.16$ of oscillations in Fig.~\ref{fig:five} (a). The respective temporal period of the spin flip is characterized by $T=2\pi/\kappa v_{\rm out}$, where $v_{\rm out}$ is the average velocity of the outward motion of the spin-flipping soliton. Since  $v_{\rm out}\approx 0.409$ in Fig.~\ref{fig:five} (a) this estimate gives $T\approx 15.36$ which is close to the exact numerical period $15.69$. Thus after the collision the outgoing breathing spinors adiabatically follow the helicoidal SO coupling.

A striking effect is the acceleration of the spinors with flipping spin, especially in the interaction shown in Fig.~\ref{fig:five} (c), where both outgoing (repelled from each other) solitons have higher average velocities than the initial pulses. This is in contrast to slowdown of the left soliton in the interaction shown in Fig.~\ref{fig:five} (b). Understanding of this phenomenon (in our system conserving the energy) resides in the dependences shown in Fig.~\ref{fig:four} (b) and (d). Indeed the chosen initial velocity $v=0.14$ is close to the extrema of the dependencies $N_\pm(v)$. Thus even small exchange of atoms between the two components occurring upon inelastic collision can accelerate or decelerate both solitons, depending on the details of the atom exchange. Additionally, the energy of the SO coupling, $\alpha\int\bPsi^\dagger\sigma_1p\bPsi dx$ is not sign-definite, and excitation of spinor rotation may lead to increase of the kinetic energy $\frac 12 \int\bPsi^\dagger p^2\bPsi dx$ of the soliton (the total energy being conserved).

To conclude, we reported the existence and stability of families of steadily moving solitons in a helicoidal gauge potential. In the absence of Zeeman splitting such solitons constitute four-parametric families, carry nonzero spin, and interact elastically similarly to solitons in integrable systems. Solitons moving with a constant velocity exist also in the presence of the Zeeman splitting. The latter, however, results in non-elastic collisions of solitons and in excitations of spinor breathers, characterized by varying $z$-component of the spin. {\color{black} Each soliton can be viewed as a quasi-particle carrying a spin degree of freedom (by analogy with an electron) whose properties can be controlled by SO coupling, that makes them suitable objects for developing spintronics in nonlinear settings. }

\acknowledgments
VVK is indebted to Prof. E. Ya. Sherman for stimulating discussions originating this work and for useful comments. YVK acknowledges support from the Severo Ochoa program (SEV-2015-0522) of the Government of Spain, from Fundacio Cellex, Generalitat de Catalunya and CERCA. VVK was supported by the FCT (Portugal) grants UID/FIS/00618/2013.

\end{document}